\documentstyle[12pt,epsfig]{article}
\newcommand{\be}{\begin{equation}}
\newcommand{\ee}{\end{equation}}
\newcommand{\beq}{\begin{equation}}
\newcommand{\eeq}{\end{equation}}
\newcommand{\ba}{\begin{eqnarray}}
\newcommand{\ea}{\end{eqnarray}}
\newcommand{\bea}{\begin{eqnarray}}
\newcommand{\eea}{\end{eqnarray}}

\begin{document}
\baselineskip=15.5pt \pagestyle{plain} \setcounter{page}{1}
\begin{titlepage}

\begin{center}

{\LARGE Plasma conductivity at finite coupling} \vskip .3cm

\vskip 1.cm

{\large {Babiker Hassanain\footnote{\tt babiker@thphys.ox.ac.uk}$
^{, \, a}$ and Martin Schvellinger\footnote{\tt
martin@fisica.unlp.edu.ar}$^{, \, b}$} }

\vskip 1.cm

{\it $^a$ The Rudolf Peierls Centre for Theoretical Physics, \\
Department of Physics, University of Oxford. \\ 1 Keble Road,
Oxford, OX1 3NP, UK.} \\

\vskip 0.5 cm

{\it $^b$ IFLP-CCT-La Plata, CONICET and \\
Departamento  de F\'{\i}sica, Universidad Nacional de La Plata.
\\ Calle 49 y 115, C.C. 67, (1900) La Plata,  \\ Buenos Aires,
Argentina.} \\

\vspace{1.cm}

{\bf Abstract}

\end{center}

By taking into account the full ${\cal {O}}(\alpha'^3)$ type IIB
string theory corrections to the supergravity action, we compute the
leading finite 't Hooft coupling ${\cal {O}}(\lambda^{-3/2})$
corrections to the conductivity of strongly-coupled $SU(N)$ ${\cal
{N}}=4$ supersymmetric Yang-Mills plasma in the large $N$ limit. We
find that the conductivity is {\emph{enhanced}} by the corrections,
in agreement with the trend expected from previous perturbative
weak-coupling computations.

\noindent

\end{titlepage}

\newpage


\vfill

\section{Introduction}

The gauge/string duality \cite{Maldacena:1997re,Gubser:1998bc,
Witten:1998qj} has become an essential framework for the study of
strongly-coupled systems, in which perturbative quantum field
theory techniques cannot be successfully applied.
Particularly, during the last decade, and with remarkable results,
gauge/string duality techniques have been applied to study
properties of the deconfined quark-gluon plasma (QGP), which is
produced as a result of heavy ion collisions at RHIC and the LHC.
One may infer that the reason for the relative success of the
gauge/string duality applied to the QGP is that the QGP produced in
those experiments is in fact strongly-coupled. Some references
discussing the phenomenology of the QGP are
\cite{Shuryak:2003xe,Gyulassy:2004zy,Muller:2007rs,
CasalderreySolana:2007zz,Shuryak:2008eq,Heinz:2008tv,Iancu:2008sp}.
For a summary of the results of the Pb-Pb ion collisions carried out
recently at the LHC, see \cite{LHC-heavy-ion}, and for an
interesting discussion of the predictions of heavy ion collisions at
the LHC we refer the reader to \cite{Abreu:2007kv}.

While the work involving the gauge/string duality in the top-down
sense mainly deals with highly symmetric field theories, the real
world at low energies is governed by QCD. One of the most
challenging aspects of studying the experimentally-produced QGP is
how to deform the gravity backgrounds available in string theory in
order to break some of the symmetries of the dual field theories,
thus bringing them closer to QCD. Another approach is to improve the
AdS/CFT results for highly symmetric theories, in order to
understand those theories better despite their differences with QCD,
in the hope that one can extract universal properties, or make
statements that apply to a wide class of strongly-coupled field
theories. In this work, we adopt this approach, focussing our study
on the hydrodynamics, and in particular charge transport, in
${\cal{N}}=4$ SYM theory at finite yet strong 't Hooft coupling.

The hydrodynamic regime of the ${\cal{N}}=4$ SYM plasma has been
studied extensively over the past decade. Using the rules developed
in references \cite{Son:2002sd,Policastro:2002se}, the two-point
correlators for the energy-momentum tensor $T_{\mu\nu}$, and the
$R$-charge vector currents $J_\mu$ in the low-momentum
(hydrodynamic) regime have been computed. The transport coefficients
of both energy-momentum and charge were extracted from the two-point
functions, yielding the viscosity
\cite{Policastro:2001yc,Kovtun:2003wp,Kovtun:2004de} and the
$R$-charge conductivity
\cite{Policastro:2002se,Kovtun:2003wp,Teaney:2006nc,Myers:2007we,CaronHuot:2006te}.
The work done in these references pertained to the large $N$ limit
(where $N$ is the rank of the gauge group, {\it i.e.} for an
infinite number of colour degrees of freedom), and infinite 't Hooft
coupling ($\lambda$). Therefore, the gravitational dual in which the
work was carried out was given by zeroth order solutions of type IIB
supergravity, {\it i.e.} the gravity backgrounds were computed to
zeroth order in $\alpha'$, consistent with the assertion that the
$\lambda \to \infty$ limit in the field theory corresponds to the
$\alpha' \to 0$ limit on the gravity side. Precisely, the solutions
obtained from the minimal quadratic type IIB supergravity action,
correct to ${\cal{O}}(\alpha'^0)$, are perturbed in the directions
which source the relevant boundary operator, obtaining an
${\cal{O}}(\alpha'^0)$ action for the supergravity perturbation. The
equations of motion derived from this action are solved, thus
obtaining the correlators of operators of ${\cal{N}}=4$ SYM theory
at infinite 't Hooft coupling. In the case of ${\cal {N}}=4$ SYM
plasma, the ${\cal{O}}(\alpha'^0)$ ten-dimensional supergravity
background is the product of an AdS$_5$-Schwarzschild black hole and
a five-sphere, with a constant Ramond-Ramond five-form field
strength and a constant dilaton. All ${\cal {N}}=4$ SYM correlators
obtained by perturbing this background are correct at infinite 't
Hooft coupling.

A natural question to then ask is the following: what are leading
order finite-coupling corrections to the hydrodynamic transport
coefficients in ${\cal {N}}=4$ SYM plasma? These turn out to arise
at order $\lambda^{-3/2}$, and to compute them from supergravity one
must include ${\cal{O}}(\alpha'^3)$ corrections to the minimal
supergravity action. This was carried out for the shear viscosity
and the mass-density diffusion constant in
\cite{Buchel:2004di,Benincasa:2005qc,Buchel:2008sh,Myers:2008yi,
Buchel:2008ae,Myers:2009ij,Cremonini:2009sy}. As mentioned above,
momentum-transport is governed by correlators of the energy-momentum
tensor $T_{\mu\nu}$, the dual field of which is $h_{\mu\nu}$, which
is the graviton of the AdS theory, where $\mu,\nu$ are both in the
AdS factor of the bulk geometry.

In this article, we wish to compute the ${\cal{O}}(\lambda^{-3/2})$
corrections to the electrical conductivity of the plasma, obtained
from the correlators of $J_\mu$, the dual field to which is $A_\mu$,
one of the $SO(6)$ gauge fields in the five-dimensional AdS theory.
The ten-dimensional parents of these gauge fields are $h_{\mu a}$,
where $\mu$ is in the AdS factor, and $a$ is in the $S^5$. To obtain
the corrections to the correlators of $J_\mu$, we must therefore
compute the equations of motion of $h_{\mu a}$ starting from the
ten-dimensional type IIB supergravity action {\it{plus}} the
${\mathcal{O}}(\alpha'^3)$ string theory corrections. The schematic
form of the ${\mathcal{O}}(\alpha'^3)$ corrections is $C^4+ C^3
{\mathcal{T}}+C^2 {\mathcal{T}}^2 + C {\mathcal{T}}^3 +
{\mathcal{T}}^4$, where $C$ is the ten-dimensional Weyl tensor,
while the rank-6 tensor ${\mathcal{T}}$ is defined in terms of the
Ramond-Ramond five-form field strength and its covariant derivative.
The big difference between this computation and the equivalent one
for the shear viscosity is that, as shown in
\cite{Buchel:2004di,Benincasa:2005qc,Buchel:2008sh,Myers:2008yi,
Buchel:2008ae,Myers:2009ij,Cremonini:2009sy}, the only operator that
affects the tensor fluctuations $h_{\mu\nu}$ (and hence the shear
viscosity) is the operator $C^4$, which only involves the metric in
ten dimensions. For the fluctuations $h_{\mu a}$ which govern
charge-transport, on the other hand, we must include the full set of
${\mathcal{O}}(\alpha'^3)$ operators listed above, which makes the
problem much more challenging. In our recent work
\cite{Hassanain:2010fv}, we considered this problem from a general
viewpoint, without performing the dimensional reduction
explicitly{\footnote{Previously we also have studied vector
fluctuations on the ${\cal {O}}(\alpha'^3)$-corrected metric for
short distances compared to the inverse of the plasma equilibrium
temperature \cite{Hassanain:2009xw}.}}$^{,}${\footnote{Corrections
to the holographic conductivity from certain higher-dimensional
operators were considered in \cite{Ritz:2008kh,Myers:2009ij}.}. In
the work we present in this article, we compute the exact and full
correction to the electrical conductivity at order
${\cal{O}}(\lambda^{-3/2})$. We find that the conductivity is
enhanced by the corrections, giving:
\be
\sigma(\lambda) = \sigma_{\infty} \, \left(1 + \frac{\zeta(3)}{8} \,
\frac{14993}{9} \, \lambda^{-3/2} \right) \, ,
\ee
where $\sigma_{\infty}$ is the conductivity at infinite 't Hooft
coupling, and $\zeta(3) \sim 1.202$ is the Riemann Zeta function. In
the next section, we briefly discuss the interpretation of this
quantity as the electrical conductivity of the plasma, before
explaining in detail how we obtained this result.

\section{Plasma Conductivity}

We would like to compute the electrical conductivity of the $SU(N)$
${\cal{N}}=4$ SYM plasma in the large $N$ limit. We mean the
following \cite{CaronHuot:2006te}: the ${\cal{N}}=4$ SYM theory has
a global $SU(4)$ $R$-symmetry group. Focus on a $U(1)$ subgroup of
this $SU(4)$ symmetry, which, being global, does not come with any
gauge fields. Then, let us gauge this $U(1)$ group, and couple the
newly-introduced $U(1)$ gauge-field minimally to the SYM Lagrangian
with coupling $e$, in the usual way. We also make the $U(1)$
gauge-field dynamical by adding its kinetic term to the action. We
call this theory SYM-EM, following the notation of
\cite{CaronHuot:2006te}. The current which couples to the $U(1)$
gauge field is $J^{em}_\mu$, and to leading order in the coupling
$e$, is given by $J^3_{\mu}$ (where the superscript $3$ simply
signifies that this is the $R$-symmetry current in the un-gauged
$U(1)$ direction) {\it{plus}} some terms which are subleading in
$e$. Therefore, to leading order in $e$ and to full non-perturbative
order in $\lambda$ (the 't Hooft coupling of the ${\cal{N}}=4$ SYM
theory), the two-point function of $J^{em}_{\mu}$ is given by the
two-point function of the $R$-symmetry currents $J^3_\mu$ calculated
entirely in the ${\cal{N}}=4$ SYM theory. The problem thus
simplifies to computing the two-point function of the $R$-symmetry
currents $J_\mu$.

The quantity we are interested in is the retarded correlator of
$R$-symmetry currents at non-zero frequency $\omega$ and vanishing
three-momentum $\vec{q}$, defined by
\be
R_{\mu\nu}(\omega, \vec{q} = 0) = -i \int d^4x \, e^{-i \omega t} \,
\Theta(t) \, <[J_\mu(x), J_\nu(0)]> \, , \label{Rmunu}
\ee
where $\Theta(t)$ is the usual Heaviside function, and $J_\mu(x)$ is
the conserved current associated with the relevant $U(1)$ subgroup
of the $R$-symmetry group. The brackets denote the expectation value
considered as a thermal average over the statistical ensemble of an
${\cal {N}}=4$ SYM plasma at equilibrium temperature $T$. Setting
the  electromagnetic coupling to be $e$, the electrical conductivity
$\sigma$ of the plasma is given by
\be
\sigma = -\lim_{\omega \rightarrow 0}\textrm{Im} \frac{e^2}{\omega}
R_{xx}(\omega, \vec{q}=0) \, .
\ee
Our aim in this work is to derive the conductivity $\sigma$, working
in the holographic dual model, and including the full set of
${\cal{O}}(\alpha'^3)$ corrections to type IIB supergravity. In the
next section, we define the ten-dimensional corrected background and
describe in detail the field which is dual to the current $J_\mu$,
the vector perturbation $A_\mu$. We then derive the equations of
motion for the $A_x$ component of the $U(1)$ gauge field, and use
their solution to obtain the plasma conductivity corrected to
${\cal{O}}(\lambda^{-3/2})$.

\section{Setting the holographic dual background}

The type IIB supergravity action corrected to ${\cal{O}}(\alpha'^3)$
is given by:
\be
S_{IIB}=S_{IIB}^{0} + S_{IIB}^{\alpha'} \, ,
\ee
where $S_{IIB}^{0}$ denotes the minimal (two-derivative) type IIB
supergravity action, and $S_{IIB}^{\alpha'}$ encodes the
corrections. The minimal action $S_{IIB}^{0}$ contains the
Einstein-Hilbert action coupled to the dilaton and the Ramond-Ramond
five-form field strength
\be
S_{IIB}^{0}=\frac{1}{2 \kappa_{10}^2}\int \, d^{10}x\,
\sqrt{-G}\left[R_{10}-\frac{1}{2}\left(\partial\phi
\right)^2-\frac{1}{4.5!}\left(F_5\right)^2 \right] \,
.\label{action-10D}
\ee
The leading 't Hooft coupling corrections, on the other hand,
are accounted for by the following schematic action
\cite{Paulos:2008tn,Myers:2008yi}
\be\label{10DWeyl}
S_{IIB}^{\alpha'}=\frac{R^6}{2 \kappa_{10}^2}\int \, d^{10}x\,
\sqrt{-G}\left[ \, \gamma e^{-\frac{3}{2}\phi} \left(C^4 +
C^3{\mathcal{T}}+C^2{\mathcal{T}}^2+C{\mathcal{T}}^3+{\mathcal{T}}^4
\right)\right] \, , \label{10d-corrected-action}
\ee
where $\gamma$ provides the dependence on the 't Hooft coupling
$\lambda$ through the definition $\gamma \equiv \frac{1}{8} \,
\zeta(3) \, (\alpha'/R^2)^{3}$, with $R^4 = 4 \pi g_s N \alpha'^2$
and $\zeta$ stands for the Riemann zeta function. Setting $\lambda =
g_{YM}^2 N \equiv 4 \pi g_s N$, we get $\gamma = \frac{1}{8} \,
\zeta(3) \, \frac{1}{\lambda^{3/2}}$.

The $C^4$ term is a dimension-eight operator, defined by the
following contractions
\be
C^4=C^{hmnk} \, C_{pmnq} \, C_h^{\,\,\,rsp} \, C^{q}_{\,\,\,rsk} +
\frac{1}{2} \, C^{hkmn} \, C_{pqmn} \, C_h^{\,\,\, rsp} \,
C^q_{\,\,\, rsk} \, ,
\ee
where $C^{q}_{\,\,\, rsk}$ is the Weyl tensor. We define the tensor
${\cal {T}}$ as:
\begin{equation}
{\cal {T}}_{abcdef}= i\nabla_a
F^{+}_{bcdef}+\frac{1}{16}\left(F^{+}_{abcmn}F^{+}_{def}{}^{mn}-3
F^{+}_{abfmn}F^{+}_{dec}{}^{mn}\right) \, , \label{T-tensor}
\end{equation}
where the RHS must be antisymmetrized in $[a,b,c]$ and $[d,e,f]$ and
symmetrized with respect to interchange of $abc \leftrightarrow def$
\cite{Paulos:2008tn}, and in addition we have
\be
F^{+} = \frac{1}{2} (1+\ast) F_5 \, .
\ee
The dual to ${\cal{N}}=4$ $SU(N)$ SYM plasma at infinite 't Hooft
coupling and infinite $N$ is the following maximally-symmetric
solution of the equations of motion of the minimal action
$S_{IIB}^{0}$: an AdS$_5$-Schwarzschild black hole multiplied by a
five-sphere. The five-form field strength is the volume form on the
sphere $\epsilon$, with $N$ units of flux through the sphere. This
is a solution of the equations of motion arising from the action
$S_{IIB}^{0}$ of Eq.(\ref{action-10D}). The current operator of the
SYM theory $J_\mu(x)$ is dual to the $s$-wave mode of the vectorial
fluctuation on this background, and we shall denote it by $A_\mu$.
In order to obtain the Lagrangian for the vectorial perturbation in
this background, we have to construct a consistent perturbed {\it
Ansatz} for both the metric and the Ramond-Ramond five-form field
strength, such that a proper $U(1)$ subgroup of the $R$-symmetry
group is obtained \cite{Cvetic:1999xp,Chamblin}. As shown in the
previous references, the consistent perturbation {\it Ansatz} yields
the minimal $U(1)$ gauge field kinetic term in the
AdS$_5$-Schwarzschild black-hole geometry. Therefore, by studying
the bulk solutions of the Maxwell equations in the
AdS$_5$-Schwarzschild black hole with certain boundary conditions,
we can obtain the retarded correlation functions \cite{Son:2002sd,
Policastro:2002se,CaronHuot:2006te} of the operator $J_\mu(x)$. We
wish to carry out this program using the fully-corrected action
$S_{IIB}$. As pointed out in the introduction, this program has
already been carried out for the tensor fluctuations $h_{\mu\nu}$
dual to the energy momentum tensor $T_{\mu\nu}$
\cite{Buchel:2004di,Buchel:2008sh,Sinha:2009ev}, yielding
 the viscosity and mass-diffusion constant of the SYM plasma.

To begin the computation of the ${\mathcal{O}}(\alpha'^3)$ equations
of motion of the gauge field $A_\mu$ dual to the current $J_\mu$, we
must first describe the effects of $S_{IIB}^{\alpha'}$ on the
geometry of the gravitational dual to the SYM plasma. Firstly, we
note that the higher curvature corrections do not modify the metric
at zero temperature \cite{Banks:1998nr}. This is simply the
statement that the SYM theory is fully conformal at all orders at
zero temperature, and so the gravitational dual must necessarily
reflect this preserved conformality. The situation at finite
temperature is rather different. In references
\cite{Gubser:1998nz,Pawelczyk:1998pb}, the effect of the string
theory leading corrections to the metric were investigated,
focussing upon the study of their corrections to thermodynamic
quantities of the five-dimensional AdS-Schwarzschild black hole. The
corrections to the metric were revisited in references
\cite{deHaro:2002vk,deHaro:2003zd,Peeters:2003pv}. Remarkably,
Myers, Paulos and Sinha \cite{Myers:2008yi} have shown that the
metric itself is only corrected by $C^4$, a consequence of the fact
that the tensor ${\mathcal{T}}$ vanishes on the uncorrected
supergravity solution. The corrected metric is given by
\cite{Gubser:1998nz,Pawelczyk:1998pb,deHaro:2003zd}
\be
ds^2 = \left(\frac{r_0}{R}\right)^2\frac{1}{u} \, \left(-f(u) \,
K^2(u) \, dt^2 + d\vec{x}^2\right) + \frac{R^2}{4 u^2 f(u)} \,
P^2(u) \, du^2 + R^2 L^2(u) \, d\Omega_5^2 \, ,\label{proper-metric}
\ee
where $f(u)=1-u^2$ and $R$ is the radius of the AdS$_5$ and the
five-sphere. The AdS-boundary is at $u=0$ and the black hole horizon
is at $u=1$. For the AdS$_5$ coordinates we use indices $m$, where
$m= \{(\mu=0, 1, 2, 3),5\}$. We have
\be
K(u) = \exp{[\gamma \, (a(u) + 4b(u))]} \, , \quad P(u) =
\exp{[\gamma \, b(u)]} \, , \quad L(u) =  \exp{[\gamma \, c(u)]} \,
,
\ee
where the exponents are given by:
\ba
a(u) &=& -\frac{1625}{8} \, u^2 - 175 \, u^4 + \frac{10005}{16} \,
u^6 \, , \nonumber \\
b(u) &=& \frac{325}{8} \, u^2 + \frac{1075}{32} \, u^4
- \frac{4835}{32} \, u^6 \, , \nonumber \\
c(u) &=& \frac{15}{32} \, (1+u^2) \, u^4 \, .
\ea
Finally, we have the following expression for the extremality
parameter $r_0$:
\be
r_0 = \frac{\pi T R^2}{(1+\frac{265}{16} \gamma)} \,,
\ee
where $T$ is identified as the {\emph{physical}} equilibrium
temperature of the plasma. Having obtained the corrected metric, the
next step is to deduce the appropriate perturbation {\it
Ans${\ddot{a}}$tze} for the vectorial fluctuations $A_\mu$ of the
corrected supergravity background. Notice that the vector
perturbation enters the perturbed metric {\emph{in addition}} to the
perturbed $F_5$ solution, which means that all the operators inside
$S_{IIB}^{\alpha'} $ can influence the calculation. The plan is to
formulate the perturbation {\it Ans${\ddot{a}}$tze}, correct to
linear order in $\gamma$ and plug them into $S_{IIB}$. Due to the
fact that we are working to linear order in $\gamma$, we must insert
the fully corrected {\it Ans${\ddot{a}}$tze} into the $S_{IIB}^{0}$
piece, but it is sufficient to insert the ${\cal{O}}(\gamma^0)$ {\it
Ans${\ddot{a}}$tze} into $S_{IIB}^{\alpha'}$ because the latter part
of the action carries an explicit factor of $\gamma$. In the
remainder of this section, we will describe the perturbation {\it
Ans${\ddot{a}}$tze} and display the result of inserting them into
$S_{IIB}^{0}$. The insertion of the perturbation {\it
Ans${\ddot{a}}$tze} into $S_{IIB}^{\alpha'}$ will be described in
the next section.

The metric {\it Ansatz} reads as follows
\ba
ds^2&=&\left[ g_{mn}+ \frac{4}{3} R^2 L(u)^2\, A_m A_n \right]
\,dx^m dx^n + R^2 L(u)^2 \, d\Omega_5^2  + \frac{4}{\sqrt{3}} R^2
L(u)^2 \nonumber \\
&& \times \left(\sin^2y_1 \, dy_3 + \cos^2 y_1 \, \sin^2 y_2 \, dy_4
+ \cos^2 y_1 \, \cos^2 y_2 \, dy_5 \right) \, A_m \, dx^m \,
\label{metric-ansatz} ,
\ea
where $d\Omega_5^2$ is metric of the unit five-sphere given by
\be
d\Omega_5^2  = dy_1^2 + \cos^2 y_1 \, dy_2^2  + \sin^2 y_1 \, dy_3^2
+ \cos^2 y_1 \, \sin^2 y_2 \, dy_4^2  + \cos^2 y_1 \ \cos^2y_2 \,
dy_5^2 \, .\nonumber
\ee
Since we are only interested in the terms which are
quadratic in the gauge-field perturbation we can write the $F_5$
{\it Ansatz} as follows
\be
F_5 = -\frac{4}{R} \overline{\epsilon} + \frac{R^3 L(u)^3}{\sqrt{3}}
\, \left( \sum_{i=1}^3 d\mu_i^2 \wedge d\phi_i \right) \wedge
\overline{\ast} F_2 \, , \label{F5ansatzCorrected}
\ee
where $F_2 = dA$ is the Abelian field strength and
$\overline{\epsilon}$ is a deformation of the volume form of the
metric of the AdS$_5$-Schwarzschild black hole. We stress that we
are not interested in the part of $F_5$ which does not contain the
vector perturbations, as we are only concerned with the quadratic
action of $A_\mu$. The Hodge dual $\ast$ is taken with respect to
the ten-dimensional metric, while $\overline{\ast}$ denotes the
Hodge dual with respect to the five-dimensional metric piece of the
black hole. In addition, we have the usual definitions for the
coordinates on the $S^5$
\ba
&& \mu_1 = \sin y_1 \, ,\,\,\,\,\,\,\,\,\,\,\,\,\,\,\,\,\,\,\, \mu_2
= \cos y_1 \, \sin y_2 \, , \,\,\,\,\,\,\,\,\,\,\,\,\,\,\,\,\,\,\,
\mu_3 = \cos y_1 \, \cos y_2 \, , \nonumber \\
&& \phi_1 = y_3 \, ,
\,\,\,\,\,\,\,\,\,\,\,\,\,\,\,\,\,\,\,\,\,\,\,\,\, \phi_2 = y_4 \, ,
\, \,\,\,\,\,\,\,\,\,\,\,\,\,\,\,\,\,\,\,\,\,\,\,\,\,\,\,
\,\,\,\,\,\,\,\,\,\,\,\,\,\,\, \phi_3 = y_5 \, .
\ea
Inserting these {\it Ans${\ddot{a}}$tze} into Eq.(\ref{action-10D}),
and discarding all the higher (massive) Kaluza-Klein harmonics of
the five-sphere, we get the following action for the zero-mode
Abelian gauge field $A_m$
\be
S_{IIB}^{SUGRA} = -\frac{{\tilde {N}}^2}{64 \pi^2 R} \int d^4x \, du
\, \sqrt{-g} \, L^7(u) \, g^{mp} \, g^{nq} \, F_{mn} \, F_{pq} \, .
\label{Fsquared}
\ee
In the previous equation, the Abelian field strength is defined as
$F_{mn}=\partial_m A_n -\partial_n A_m$, the partial derivatives are
$\partial_m =
\partial/\partial x^m$, while $x^m=(t, \vec{x}, u)$, where $t$ and
$\vec{x}=(x_1, x_2, x_3)$ are the Minkowski four-dimensional
spacetime coordinates, and $g \equiv \textrm{det} (g_{mn})$, where
$g_{mn}$ is the metric of AdS$_5$-Schwarzschild black hole. The
factor of $L(u)^7$ arises straightforwardly from the dimensional
reduction \cite{Kovtun:2003wp}, and the volume of the five-sphere
has been included in ${\tilde {N}}$.

The next step is to obtain the effect of the eight-derivative
corrections of Eq.(\ref{10DWeyl}). As in \cite{Hassanain:2010fv}, it
is sufficient to use the uncorrected {\it Ans${\ddot{a}}$tze} at
this point, {\it i.e.} Eqs.(\ref{metric-ansatz}) and
(\ref{F5ansatzCorrected}) in the limit $\gamma \to 0$, so taking
$L(u),K(u),P(u)\to 1$ and $\overline{\epsilon}\to \epsilon$. We
carry this out in the next section.

\section{'t Hooft corrections to the $R$-charge conductivity}

We here insert the {\it Ans${\ddot{a}}$tze} in
Eqs.(\ref{metric-ansatz}) and (\ref{F5ansatzCorrected}) (with
$\gamma \to 0$) into $S_{IIB}^{\alpha'}$. We begin by explicitly
writing all the various operators comprising $S_{IIB}^{\alpha'}$.
For this purpose it is convenient to use the definitions given by
Paulos \cite{Paulos:2008tn} which can be explicitly written from
Eq.(\ref{10d-corrected-action}) as
\ba
S_{IIB}^{\alpha'} &=& \frac{R^6}{2 \kappa_{10}^2} \, \int \, d^{10}x
\, \sqrt{-G} \, [\gamma e^{-\frac{3}{2}\phi} \, (C^4 +
C^3{\mathcal{T}}+C^2{\mathcal{T}}^2+C{\mathcal{T}}^3+{\mathcal{T}}^4)]
\nonumber \\
&\equiv& \frac{1}{86016} \, \frac{R^6}{2 \kappa_{10}^2} \, \sum_i
n_i \, \int \, d^{10}x \, \sqrt{-G} \, [\gamma e^{-\frac{3}{2}\phi}
\, M_i ] \, ,
\ea
where the coefficients $n_i$ are found in \cite{Paulos:2008tn}.
The first important point to keep in mind is that we are only
interested in terms quadratic in the gauge field $A_\mu$. We
therefore expand the tensors $C$ and ${\cal {T}}$ as follows:
$C=C_0+C_1+C_2$, and ${\cal {T}} = {\cal {T}}_0 + {\cal {T}}_1 +
{\cal {T}}_2$, where the subindex labels the number of times that
the Abelian gauge field occurs. The second important point to note
is that the tensor ${\cal {T}}_0$ vanishes for the background
considered here (and for any direct-product background which
contains a five-dimensional Einstein manifold as the internal
space), as proved by \cite{Myers:2008yi}. Therefore, we may write
${\cal {T}} = {\cal {T}}_1 + {\cal {T}}_2$. This immediately means
that the terms $C{\mathcal{T}}^3$ and ${\mathcal{T}}^4$ cannot
contribute to the quadratic action for the gauge field $A_\mu$ and
so we discard them in what follows.

For $C^4$ there are two contributions which can be written as:
\be
C^4 = - \frac{43008}{86016}  C_{abcd} C_{abef} C_{cegh} C_{dgfh} +
C_{abcd} C_{aecf} C_{bgeh} C_{dgfh} \, ,
\ee
where repeated indices mean usual Lorentz contractions. The
contributions from this term can be schematically written as $C_0^3
C_2$ and $C_0^2 C_1^2$.

Next we consider terms of the form $C^3 \, {\cal {T}}$
\be
C^3 {\cal {T}} = \frac{3}{2} \, C_{abcd} C_{aefg} C_{bfhi} {\cal
{T}}_{cdeghi} \, .
\ee
The possible contributions from these terms are of the form $C_0^2
C_1{\cal {T}}_1$ and $C_0^3 {\cal{T}}_2$. We have explicitly checked
that the $C_0^3 {\cal{T}}_2$ term is zero, so that the only
contribution here is $C_0^2 C_1{\cal{T}}_1$.

For the operators $C^2 {\cal {T}}^2$, there are a few
contractions:
\ba
C^2 {\cal {T}}^2 =  \frac{1}{86016} &\times & \left( 30240 \,
C_{abcd} C_{abce} {\cal {T}}_{dfghij} {\cal {T}}_{efhgij} +  7392 \,
C_{abcd} C_{abef}
{\cal {T}}_{cdghij} {\cal {T}}_{efghij} \right. \nonumber \\
&& \left. - 4032 \, C_{abcd} C_{aecf} {\cal {T}}_{beghij} {\cal
{T}}_{dfghij} - 4032 \, C_{abcd} C_{aecf}
{\cal {T}}_{bghdij} {\cal {T}}_{eghfij} \right. \nonumber \\
&& \left. - 118272 \, C_{abcd} C_{aefg} {\cal {T}}_{bcehij} {\cal
{T}}_{dfhgij} -  26880 \, C_{abcd} C_{aefg}
{\cal {T}}_{bcehij} {\cal {T}}_{dhifgj} \right. \nonumber \\%
&& \left. +  112896 \, C_{abcd} C_{aefg} {\cal {T}}_{bcfhij} {\cal
{T}}_{dehgij} - 96768 \, C_{abcd} C_{aefg} {\cal {T}}_{bcheij} {\cal
{T}}_{dfhgij} \right). \nonumber \\
&&
\ea
The contributions here are of the form $C_0^2 {\cal {T}}_1^2$.

What we must now do is clear: compute the ten-dimensional Weyl
tensor $C$ to quadratic order in the gauge field $A_x$, and compute
${\cal {T}}_1$. We separate the latter into ${\cal {T}}_1 = \nabla
F_5 + {\bar {\cal {T}}}$, where
\be
({\nabla F_5})_{abcdef}= i\nabla_a F^{+}_{bcdef} \, ,
\ee
and a second piece which does not contain covariant derivatives
\be
\bar {\cal {T}}_{abcdef}  =
\frac{1}{16}\left(F^{+}_{abcmn}F^{+}_{def}{}^{mn}-3
F^{+}_{abfmn}F^{+}_{dec}{}^{mn}\right) \, .
\ee
We can recast the definition of $F^{+}$ above as a sum of an
electric and magnetic components $F^{+} = F_{(e)} + F_{(m)}$. For
the electric part we have
\be
F_{(e)} = -\frac{4}{R} \epsilon + \frac{R^3}{\sqrt{3}} \, \left(
\sum_{i=1}^3 d\mu_i^2 \wedge d\phi_i \right) \wedge \overline{\ast}
F_2 \, ,
\ee
where the $\overline{\ast}$ indicates the Hodge dual with respect to
the AdS$_5$-Schwarzschild black hole metric. It is
convenient to split the electric part into a background piece plus a
fluctuation piece: $F_{(e)}=F_{(e)}^{(0)}+F_{(e)}^{(f)}$, and
similarly for the magnetic terms. Then, in components, we can write:
\be
(F_{(e)}^{(0)})_{\mu\nu\rho\sigma\delta} = - \frac{4}{R} \,
\sqrt{-g} \, \epsilon_{\mu\nu\rho\sigma\delta} \, ,
\ee
where as before $g$ is the determinant of the AdS
piece of the metric. The Hodge dual of the previous equation gives
\be
(F_{(m)}^{(0)})_{abcde} = - \frac{4}{R} \, R^5 \, \sqrt{\det{S^5}}
\epsilon_{abcde} \, .
\ee
For the conductivity, we are free to restrict ourselves to the $A_x(u)$
component of the Abelian field. Thus, $F_2$ has only one component,
namely $F_{ux}(u)$. We write $F=dA=\frac{1}{2!} F_{\mu\nu}/\sqrt{3}
\, dx^\mu \wedge dx^\nu$, obtaining for the fluctuation of the
electric part:
\be
(F_{(e)}^{(f)})_{y_iy_jtyz} = - \frac{R^3}{\sqrt{3}} \,
\frac{b_{ij}}{2} \, \sqrt{g} \, (2 F_{ux} G^{xx} G^{uu})
\epsilon_{y_iy_jtyz} \, ,
\ee
where the pairs $(i j)$ are $(1 3)$, $(1 4)$, $(1 5)$, $(2 4)$ and
$(2 5)$. The $b_{ij}$ functions are:
\ba
&& b_{13}=2 \sin y_1 \cos y_1 \, , \quad b_{14}= - 2 \sin^2 y_2 \sin
y_1 \cos y_1
\, , \quad b_{15}=- 2 \cos^2 y_2 \sin y_1 \cos y_1 \, , \nonumber \\
&& b_{24}= 2 \cos^2y_1 \sin y_2 \cos y_2 \, , \quad b_{25}= - 2
\cos^2 y_1 \sin y_2 \cos y_2 \, .
\ea
For the fluctuations of the magnetic part we have
\ba
F_{(m)}^{(f)} &=& \sqrt{-G_{10}} \, {\tilde F}(u) \, G^{tt} \,
G^{yy} \, G^{zz}
\times \nonumber \\
&& (m_{13} \epsilon_{uxy_2y_4y_5}+m_{14}
\epsilon_{uxy_2y_3y_5}+m_{15} \epsilon_{uxy_2y_3y_4}+m_{24}
\epsilon_{uxy_1y_3y_5}+m_{25} \epsilon_{uxy_1y_3y_4}) \, ,
\ea
where $G_{10}$ is the determinant of the full ten-dimensional metric and
for conciseness we have defined
\be
{\tilde F}(u) = - \frac{R^3}{\sqrt{3}} \, \frac{1}{2} \, \sqrt{-g}
\, (2 F_{ux} G^{xx} G^{uu}) \, .
\ee
The functions $m_{ij}$ are given by
\ba
&& m_{13}= - \frac{4}{R^4} \sin(2 y_2) \cos^4 y_1 \, , \quad m_{14}=
- \frac{8}{R^4} \sin^2 y_1 \cos^2 y_1 \sin y_2 \cos y_2 \, ,
\nonumber \\
& &
m_{15}= \frac{8}{R^4} \sin^2 y_1 \cos^2 y_1 \sin y_2 \cos y_2 \, , \nonumber \\
&& m_{24}= - \frac{8}{R^4} \cos^2 y_2 \sin y_1 \cos y_1 \, , \quad
m_{25}= - \frac{8}{R^4} \cos y_1 \sin y_1 \sin^2 y_2 \, .
\ea
We are now in a position to put all the ingredients together.
We plug the {\it Ansatz} into $S_{IIB}^{\alpha'}$, multiply
by the determinant of the metric, then integrate out the
coordinates of the five-sphere. Setting $f(u)=1-u^2$ as above, the
result for the covariant derivative piece (the $C^2 (\nabla F_5)^2$
arising from $C^2 {\cal{T}}^2$ operator) is:
\ba
L_{C^2 (\nabla F_+)^2} &=& - \frac{u^4}{9} \, [(11839-30773 u^2 +
25278 u^4) \, A_x'^2 \nonumber \\
&&-2 \, u \, f(u) \, (9401 u^2-6229) A_x' A_x'' + 3773 \, u^2 \,
f(u)^2 \, (A_x'')^2] \, .
\ea
On the other hand, for the terms from $C^3 {\cal {T}}$, we get
\be
L_{C^3 {\cal {T}}_1} =  -\frac{112 u^4}{3} \, f(u) \, (A_x')^2 \, .
\ee
Similarly for the terms from $C^2 {\cal {T}}^2$ arising from the
$\bar{\cal{T}}$ piece of the tensor, we obtain
\be
L_{C^2 F_+^2} =  \frac{830 u^4}{3} \, f(u) \, (A_x')^2 \, .
\ee
Finally, for $C^4$ we obtain
\be
L_{C^4} = -\frac{u^4}{3} \left( 4 \, (14-67 u^2 +78 u^4) \, A_x'^2 +
4 \, u \, f(u) \, (28 - 53 u^2) \, A_x' \, A_x'' + 33 \, u^2 \,
f(u)^2 \, (A_x'')^2\right) \, .
\ee
We can therefore write the total Lagrangian coming directly from the
dimension-eight operators in $S_{IIB}^{\alpha'}$ as
\be\label{orderGamma}
L_{(C^4+{\mathcal{T}})^4} = c_{1} \, L_{C^2 (\nabla F_+)^2} + c_{2}
\, L_{C^2 F_+^2} + c_{3} \, L_{C^3 {\cal {T}}_1} + c_{4} \, L_{C^4}
\, ,
\ee
where all the coefficients $c_i=1$ and we include them for the
purpose of keeping track of the effects of every term in the final
expression for the electrical conductivity. The Lagrangian in
Eq.(\ref{orderGamma}) must be augmented by the terms coming from
$S_{IIB}^{0}$, that is, the kinetic term of Eq.(\ref{Fsquared}). Once that
term is added, we are left with the following Lagrangian, whose
equations of motion must be derived and solved:
\ba
S_{\textrm{total}} &=&- \frac{{\tilde {N}}^2 r_0^2}{16 \pi^2 R^4}
\int \frac{d^4k}{(2 \pi)^4} \int_0^1 du \, \left[(B_1+\gamma B_W)
A_k' A_{-k}' +\gamma E_W A_k'' A_{-k}''+\gamma F_W A_k'' A_{-k}'
\right]  \, , \label{AxAction} \nonumber \\
&&
\ea
where we have introduced the following Fourier transform of the field $A_x$
\be
A_{x}(t, \vec{x}, u) = \int \frac{d^4k}{(2 \pi)^4}
\, e^{-i \omega t + i q z} \, A_k(u) \, .
\ee
There are also a number of boundary terms that must be included for
this higher-derivative Lagrangian to make sense, and this is
discussed in detail in
\cite{Buchel:2004di,Hassanain:2009xw,Hassanain:2010fv}. In the
Lagrangian of Eq.(\ref{AxAction}), the coefficient $B_1$ arises
directly from the kinetic term $F^2$. The subscript $W$ indicates
that the particular coefficient comes directly from the
eight-derivative corrections. We have
\ba
B_1 &=&\frac{K(u)f(u) L^7(u)}{P(u)} \, , \nonumber \\
B_W &=&-\frac{u^4}{9} \left( 12 c_4\left[14-67 u^2 +78 u^4 \right] \right. \nonumber \\
&& \left. +c_1 \left[ 11839-30773 u^2 +
25278 u^4 \right] + \left[ 336 c_3 - 2490 c_2 \right] f(u) \right)   \, , \nonumber \\
E_W &=& -\frac{11}{9}\left[ 9 c_4 +343 c_1 \right] u^6 f(u)^2  \, , \nonumber \\
F_W &=&\frac{2}{9}u^5f(u)\left( 6 c_4 \left[ 53u^2-28 \right] +c_1
\left[ 9401 u^2 -6229 \right] \right) \, . \ea
We note that the equation of motion arising from (a more general
version of) this Lagrangian was solved in \cite{Hassanain:2010fv}
following \cite{Buchel:2004di}, so we will be very brief in what
follows. We also stress that the behaviour of the solution at the
black-hole horizon is unchanged by the finite coupling corrections
(once it is expressed in terms of the physical temperature and the
physical momentum), consistently with the findings of
\cite{Hassanain:2010fv}. The solution therefore has the following
form:
\be
A_k(u)=A_0(u)+\gamma A_1(u)\, = \,\left[1-u \right]^{-\delta} \left(
\phi_0(u)+\gamma \phi_1(u) \right)\, ,
\ee
where $\delta=i\omega/(4\pi T)$. Using this {\it Ansatz} and
following the work of \cite{Hassanain:2010fv}, the full solution of
the equations of motion to linear order in $\gamma$ and $\delta$ is:
\begin{equation}
A_k(u)=\left[1-u \right]^{-\delta} \left(\overline{C}+\delta
\left\{\overline{D}+\overline{C}\left(1 +\gamma \left[\frac{185}{4}
+2\beta \right] \right)u  \right\} \right) \, ,
\end{equation}
where $\overline{C}, \overline{D}$ drop out of the final result,
and we shall reveal $\beta$ shortly. We must obtain the on-shell
action for this solution. The form of the functions $B_W,E_W,F_W$
means that the on-shell action reduces to \cite{Hassanain:2010fv}
\be
S_{\textrm{on-shell}} =\frac{{\tilde {N}}^2 r_0^2}{16 \pi^2 R^4} \int
\frac{d^4k}{(2 \pi)^4} \, \left[ B_1 A_k'A_{-k} \right] \Big\vert_{u=0}  \, .
\label{ActionOnShell}
\ee
Evaluating the previous equation, and differentiating twice with
respect  to the boundary value of the gauge field, we obtain that
the conductivity of the large $N$ limit of strongly-coupled $SU(N)$
${\cal {N}}=4$ SYM plasma is corrected by the following factor:
\be
1+\gamma \left(\beta -10 \right) \, , \nonumber
\ee
where
\begin{equation}
\beta= \frac{12797}{9} \, c_{1} + \frac{2490}{9} \, c_{2} -
\frac{336}{9} \, c_{3} + \frac{44}{3} c_{4}  \, . \label{alpha}
\end{equation}
Setting the coefficient $c_i$ to their actual numerical value
$(=1)$, we obtain the following final expression for the
conductivity
\be\label{conductres}
\sigma(\lambda) = \sigma_{\infty} \, \left(1 + \frac{\zeta(3)}{8} \,
C \, \lambda^{-3/2} \right) \, ,
\ee
where the conductivity at infinite 't Hooft coupling is
\be
\sigma_{\infty} = e^2 \frac{N^2 T}{16 \pi} \, ,
\ee
where $e$ is the electric charge, and $C$ is given by
\be
C = \frac{14993}{9} \approx 1665.89 \, ,
\ee
For $\lambda=100$ the correction $\frac{\zeta(3)}{8} \, 14993/9
\, \lambda^{-3/2}$ gives 0.25031, which is a 25 percent
enhancement of the electrical conductivity compared
with its value at infinite 't Hooft coupling. For $\lambda=1000$,
the enhancement reduces to just under one
percent.

\section{Discussion and concluding remarks}

In this article, we computed the finite 't Hooft coupling
corrections to the conductivity of the large $N$ limit of the
strongly-coupled $SU(N)$ ${\cal {N}}=4$ SYM plasma. The corrections
start at ${\cal {O}}(\lambda^{-3/2})$ and enhance the conductivity
from its value at infinite strong coupling. In our previous work
\cite{Hassanain:2010fv}, we analysed the effect of the
${\cal{O}}(\alpha'^3)$ dimension-eight operators on the vector
fluctuations of the supergravity metric, through an examination of
the possible gauge-field operators that can arise in the AdS theory,
after integrating out the internal compact space. We found that the
most general initial set of 720 five-dimensional gauge-invariant
operators containing two powers of the gauge-field $A_\mu$ can be
reduced by symmetry to only 26 operators\footnote{Part of the
computation presented in reference \cite{Hassanain:2010fv},
concerning the reduction to the set of independent operators was
performed using the programme Cadabra \cite{Cadabra1,Cadabra2}.}.
The operators are of the schematic form $\tilde{C}^2 F_2^2$ and
$\tilde{C}^2 (\nabla F_2)^2$, where $\tilde{C}$ stands for the
AdS$_5$ Weyl tensor and $F_2$ for the gauge field strength
(containing $A_\mu$ above). One can easily show that the Lagrangian
obtained above in Eq.(\ref{AxAction}) is entirely consistent with
the operators obtained in \cite{Hassanain:2010fv}. Furthermore, in
\cite{Hassanain:2010fv} we showed that the functional behaviour of
the solution of $A_\mu$ at the black-hole horizon is unchanged,
provided the physical parameters of the ${\cal {N}}=4$ SYM theory
are used. To be more explicit, the solution of $A_\mu$ always takes
the form
\be
A_x(u) = \,\left[1-u \right]^{-\delta} \left(
\phi_0(u)+\gamma \phi_1(u) \right)\, ,\nonumber
\ee
where $\phi_{0,1}$ are regular at the horizon, and the index
$\delta$ is {\emph{always}} given by $\delta=i\omega/(4\pi T)$,
where $\omega$ and $T$ are the physical frequency of the
perturbation, and $T$ the physical temperature of the plasma. Again,
our results in this work are entirely consistent with this
observation. We view the agreement between our present work and that
of \cite{Hassanain:2010fv} as a good check on the correctness of our
result for the conductivity.

Before moving on we should also compare the result of our
calculation with the trend one expects from weak-coupling
computations of charge-transport coefficients. The electrical
conductivity of a weakly coupled ${\cal {N}}=4$ SYM plasma was
calculated in \cite{CaronHuot:2006te}, using the techniques used of
\cite{Arnold:2000dr}. The result of reference
\cite{CaronHuot:2006te} is
\be
\sigma = 1.28349 \, \frac{e^2 (N^2-1) T}{\lambda^2 [-\frac{1}{2} \ln
\lambda + a ]}\, ,
\ee
where $a$ is an ${\cal {O}}(1)$ constant which was not explicitly
evaluated in \cite{CaronHuot:2006te}. For perturbative values of the
't Hooft coupling, this expression implies that decreasing the 't
Hooft coupling should increase the conductivity. This trend is
confirmed by our result in Eq.(\ref{conductres}). The physical
reason here is that at weak-coupling the mean free path for particle
collisions in the plasma increases, leading to more efficient
charge-transport.

We now turn to a brief comparison between our results and lattice
QCD calculations. Firstly, the calculation above obviously has its
own value, without any reference to phenomenology, but as we
mentioned in the introduction, our ultimate aim is to make contact
with the real world, and make tentative statements about the QCD
plasma produced at RHIC and the LHC. Any such statement in the
context of our work must of course be taken with a set of caveats
and qualifiers, owing to the disparities between ${\cal {N}}=4$ SYM
theory and QCD. Indeed, these theories are very different at zero
temperature and weak coupling. However, for a temperature $T$ above
the QCD phase-transition temperature $T_c$ but not significantly higher,
both the QCD plasma and the ${\cal{N}}=4$ SYM plasma behave
like strongly-coupled ideal fluids. Moreover, there are some results from
lattice QCD implying that the thermodynamical properties of QCD are
reasonably well approximated by conformal dynamics in a range of
temperatures from about $2 T_c$ up to some high temperature (for a
discussion on numerical results from lattice QCD in comparison with
${\cal {N}}=4$ SYM theory, see \cite{Liu,Rajagopal} and references
therein). The closeness of the shear viscosity to entropy density
ratio observed in RHIC, and that computed from the gauge/gravity
duality, also lends support to the idea that there is a parametric
region where one can learn about the hydrodynamical properties of
QCD by studying the hydrodynamics of ${\cal{N}}=4$ SYM.

With this tentative philosophy in mind, it is possible to make
contact with QCD lattice calculations to some extent. We must take
into account that in those calculations $N=3$ and there are other
differences with respect to the large $N$ limit of ${\cal {N}}=4$
SYM plasma. There is a recent estimation of the conductivity given
by Aarts et al \cite{Aarts:2007wj}. That paper finds $\sigma \sim
0.4 \, e^2 \, T$, above the deconfinement transition temperature
$T_c$ of quenched lattice QCD. Assuming that the conductivity scales
with $N^2$, and setting $N=3$, a naive insertion of $\sigma \sim 0.4
\, e^2 \, T$ in our formula Eq.(\ref{conductres}) gives
$\lambda=34.52$. A more recent lattice calculation by Ding et al
\cite{Ding:2010ga} obtained $1/3 \, e^2 \, T\leq \sigma \leq e^2 \,
T$ from the vector current correlation function for light valence
quarks in the deconfined phase of quenched lattice QCD at $T =1.45
T_c$. Using these values in our formula Eq.(\ref{conductres}) yields
$14.39 \leq \lambda \leq 43.86$. It is worth noting that from
lattice computations at temperatures about 1.5 to 2 $T_c$, the
values of $\alpha_s$ are thought to be around 0.3 to 0.4, based on
heavy quark potentials\footnote{We thank Gert Aarts for this
comment.}. These values of $\alpha_s$ give a value $\lambda$ about
15. While this is of course a naive comparison which must be taken
in the context of the many caveats outlined above, it is nonetheless
pleasing that our value for the conductivity falls close to those
obtained from lattice QCD.

The results we present here have many interesting extensions.
Firstly, observe that the electrical conductivity is an extensive
quantity that depends on the number of degrees of freedom $N$ in the
theory, so it is not the best quantity for comparing
charge-transport in different theories. A more useful quantity for
this purpose is the charge-diffusion constant ${\cal{D}}$, which is
found to be $1/(2\pi T)$ in the large $N$ limit of ${\cal {N}}=4$
SYM plasma at infinite 't Hooft coupling \cite{Policastro:2002se}.
This quantity is obtained from the first pole of the $R_{zz}$
correlator, and we intend to report on the ${\cal
{O}}(\lambda^{-3/2})$ corrections to ${\cal{D}}$ in a forth-coming
paper \cite{Hassanain:2011}.

Another very interesting quantity that can be obtained from the
current two-point function $R_{xx}$ is the photoemission rate of the
plasma \cite{CaronHuot:2006te}. Computing the finite-coupling
corrections to this quantity would involve obtaining and solving the
equations of $A_x$ for the whole range of light-like momenta
($\omega = |\vec{q}|$). We will report on this in
\cite{Hassanain:2011}.

A final observation concerns the behaviour of the holographic
conductivity obtained here as we change the internal
compactification space of the ten-dimensional dual. We remind the
reader that the work of \cite{Myers:2008yi, Buchel:2008ae} proved
that the corrections to the shear viscosity to entropy density ratio
were independent of the internal compactification space for a large
class of holographic duals. Here we computed the corrections to the
conductivity using the simplest compactification space $S^5$. It
would be very interesting to investigate the impact of the
compactification space on the corrections computed here, and whether
the universality of momentum transport posited in
\cite{Myers:2008yi, Buchel:2008ae} is operative for charge transport
too.

~

~


\centerline{\large{\bf Acknowledgments}}

~

We thank Gert Aarts, Alex Buchel, Miguel Paulos and Jorge Russo for
correspondence and useful comments. B.H. thanks Christ Church
College, Oxford, for financial support. The work of M.S. has been
partially supported by the CONICET, the ANPCyT-FONCyT Grant
PICT-2007-00849, and the CONICET Grant PIP-2010-0396.


\newpage

\begin{thebibliography}{99}


\bibitem{Maldacena:1997re}
  J.~M.~Maldacena,
  ``The large N limit of superconformal field theories and supergravity,''
  Adv.\ Theor.\ Math.\ Phys.\  {\bf 2} (1998) 231
  [Int.\ J.\ Theor.\ Phys.\  {\bf 38} (1999) 1113]
  [arXiv:hep-th/9711200].

\bibitem{Gubser:1998bc}
  S.~S.~Gubser, I.~R.~Klebanov and A.~M.~Polyakov,
  ``Gauge theory correlators from non-critical string theory,''
  Phys.\ Lett.\  B {\bf 428} (1998) 105
  [arXiv:hep-th/9802109].

\bibitem{Witten:1998qj}
  E.~Witten,
  ``Anti-de Sitter space and holography,''
  Adv.\ Theor.\ Math.\ Phys.\  {\bf 2} (1998) 253
  [arXiv:hep-th/9802150].



\bibitem{Shuryak:2003xe}
  E.~Shuryak,
  ``Why does the quark gluon plasma at RHIC behave as a nearly ideal fluid?,''
  Prog.\ Part.\ Nucl.\ Phys.\  {\bf 53} (2004) 273
  [arXiv:hep-ph/0312227].

\bibitem{Gyulassy:2004zy}
  M.~Gyulassy and L.~McLerran,
  ``New forms of QCD matter discovered at RHIC,''
  Nucl.\ Phys.\  A {\bf 750} (2005) 30
  [arXiv:nucl-th/0405013].

\bibitem{Muller:2007rs}
  B.~Muller,
  ``From Quark-Gluon Plasma to the Perfect Liquid,''
  Acta Phys.\ Polon.\  B {\bf 38} (2007) 3705
  [arXiv:0710.3366 [nucl-th]].

\bibitem{CasalderreySolana:2007zz}
  J.~Casalderrey-Solana and C.~A.~Salgado,
  ``Introductory lectures on jet quenching in heavy ion collisions,''
  Acta Phys.\ Polon.\  B {\bf 38} (2007) 3731
  [arXiv:0712.3443 [hep-ph]].

\bibitem{Shuryak:2008eq}
  E.~Shuryak,
  ``Physics of Strongly coupled Quark-Gluon Plasma,''
  Prog.\ Part.\ Nucl.\ Phys.\  {\bf 62} (2009) 48
  [arXiv:0807.3033 [hep-ph]].

\bibitem{Heinz:2008tv}
  U.~W.~Heinz,
  ``The strongly coupled quark-gluon plasma created at RHIC,''
  J.\ Phys.\ A  {\bf 42} (2009) 214003
  [arXiv:0810.5529 [nucl-th]].

\bibitem{Iancu:2008sp}
  E.~Iancu,
  ``Partons and jets in a strongly-coupled plasma from AdS/CFT,''
  Acta Phys.\ Polon.\  B {\bf 39} (2008) 3213
  [arXiv:0812.0500 [hep-ph]].

\bibitem{LHC-heavy-ion}
  J.~Schukraft and f.~t.~A.~Collaboration,
  ``First Results from the ALICE experiment at the LHC,''
  arXiv:1103.3474 [hep-ex].

\bibitem{Abreu:2007kv}
  N.~Armesto {\it et al.},
  ``Heavy Ion Collisions at the LHC - Last Call for Predictions,''
  J.\ Phys.\ G {\bf 35} (2008) 054001
  [arXiv:0711.0974 [hep-ph]].


\bibitem{Son:2002sd}
  D.~T.~Son and A.~O.~Starinets,
  ``Minkowski-space correlators in AdS/CFT correspondence: Recipe and
     applications,''
  JHEP {\bf 0209} (2002) 042
  [arXiv:hep-th/0205051].

\bibitem{Policastro:2002se}
  G.~Policastro, D.~T.~Son and A.~O.~Starinets,
  ``From AdS/CFT correspondence to hydrodynamics,''
  JHEP {\bf 0209} (2002) 043
  [arXiv:hep-th/0205052].

\bibitem{Policastro:2001yc}
  G.~Policastro, D.~T.~Son and A.~O.~Starinets,
  ``The shear viscosity of strongly coupled N = 4 supersymmetric Yang-Mills
  plasma,''
  Phys.\ Rev.\ Lett.\  {\bf 87} (2001) 081601
  [arXiv:hep-th/0104066].

\bibitem{Kovtun:2003wp}
  P.~Kovtun, D.~T.~Son and A.~O.~Starinets,
  ``Holography and hydrodynamics: Diffusion on stretched horizons,''
  JHEP {\bf 0310} (2003) 064
  [arXiv:hep-th/0309213].

\bibitem{Kovtun:2004de}
  P.~Kovtun, D.~T.~Son and A.~O.~Starinets,
  ``Viscosity in strongly interacting quantum field theories from black hole
  physics,''
  Phys.\ Rev.\ Lett.\  {\bf 94} (2005) 111601
  [arXiv:hep-th/0405231].

\bibitem{Teaney:2006nc}
  D.~Teaney,
  ``Finite temperature spectral densities of momentum and R-charge correlators
  in N=4 Yang Mills theory,''
  Phys.\ Rev.\  D {\bf 74} (2006) 045025
  [arXiv:hep-ph/0602044].

\bibitem{Myers:2007we}
  R.~C.~Myers, A.~O.~Starinets and R.~M.~Thomson,
  ``Holographic spectral functions and diffusion constants for fundamental
  matter,''
  JHEP {\bf 0711}, 091 (2007)
  [arXiv:0706.0162 [hep-th]].

\bibitem{CaronHuot:2006te}
  S.~Caron-Huot, P.~Kovtun, G.~D.~Moore, A.~Starinets and L.~G.~Yaffe,
  ``Photon and dilepton production in supersymmetric Yang-Mills plasma,''
  JHEP {\bf 0612} (2006) 015
  [arXiv:hep-th/0607237].


\bibitem{Buchel:2004di}
  A.~Buchel, J.~T.~Liu and A.~O.~Starinets,
  ``Coupling constant dependence of the shear viscosity in N=4 supersymmetric
  Yang-Mills theory,''
  Nucl.\ Phys.\  B {\bf 707} (2005) 56
  [arXiv:hep-th/0406264].

\bibitem{Benincasa:2005qc}
  P.~Benincasa and A.~Buchel,
  ``Transport properties of N = 4 supersymmetric Yang-Mills theory at  finite
  coupling,''
  JHEP {\bf 0601} (2006) 103
  [arXiv:hep-th/0510041].

\bibitem{Buchel:2008sh}
  A.~Buchel,
  ``Resolving disagreement for eta/s in a CFT plasma at finite coupling,''
  Nucl.\ Phys.\  B {\bf 803} (2008) 166
  [arXiv:0805.2683 [hep-th]].

\bibitem{Myers:2008yi}
  R.~C.~Myers, M.~F.~Paulos and A.~Sinha,
  ``Quantum corrections to eta/s,''
  Phys.\ Rev.\  D {\bf 79} (2009) 041901
  [arXiv:0806.2156 [hep-th]].

\bibitem{Buchel:2008ae}
  A.~Buchel, R.~C.~Myers, M.~F.~Paulos and A.~Sinha,
  ``Universal holographic hydrodynamics at finite coupling,''
  Phys.\ Lett.\  B {\bf 669}, 364 (2008)
  [arXiv:0808.1837 [hep-th]].

\bibitem{Myers:2009ij}
  R.~C.~Myers, M.~F.~Paulos and A.~Sinha,
  ``Holographic Hydrodynamics with a Chemical Potential,''
  JHEP {\bf 0906} (2009) 006
  [arXiv:0903.2834 [hep-th]].

\bibitem{Cremonini:2009sy}
  S.~Cremonini, K.~Hanaki, J.~T.~Liu and P.~Szepietowski,
  ``Higher derivative effects on eta/s at finite chemical potential,''
  Phys.\ Rev.\  D {\bf 80} (2009) 025002
  [arXiv:0903.3244 [hep-th]].

\bibitem{Hassanain:2010fv}
  B.~Hassanain and M.~Schvellinger,
  ``Towards 't Hooft parameter corrections to charge transport in
  strongly-coupled plasma,''
  JHEP {\bf 1010} (2010) 068
  [arXiv:1006.5480 [hep-th]].

\bibitem{Hassanain:2009xw}
  B.~Hassanain and M.~Schvellinger,
  ``Holographic current correlators at finite coupling and scattering off a
  supersymmetric plasma,''
  JHEP {\bf 1004} (2010) 012
  [arXiv:0912.4704 [hep-th]].

\bibitem{Ritz:2008kh}
  A.~Ritz and J.~Ward,
  ``Weyl corrections to holographic conductivity,''
  Phys.\ Rev.\  D {\bf 79} (2009) 066003
  [arXiv:0811.4195 [hep-th]].

\bibitem{Paulos:2008tn}
  M.~F.~Paulos,
  ``Higher derivative terms including the Ramond-Ramond five-form,''
  JHEP {\bf 0810} (2008) 047
  [arXiv:0804.0763 [hep-th]].

\bibitem{Cvetic:1999xp}
  M.~Cvetic {\it et al.},
  ``Embedding AdS black holes in ten and eleven dimensions,''
  Nucl.\ Phys.\  B {\bf 558} (1999) 96
  [arXiv:hep-th/9903214].

\bibitem{Chamblin}
  A.~Chamblin, R.~Emparan, C.~V.~Johnson and R.~C.~Myers,
  ``Charged AdS black holes and catastrophic holography,''
  Phys.\ Rev.\  D {\bf 60} (1999) 064018
  [arXiv:hep-th/9902170].

\bibitem{Sinha:2009ev}
  A.~Sinha and R.~C.~Myers,
  ``The viscosity bound in string theory,''
  Nucl.\ Phys.\  A {\bf 830}, 295C (2009)
  [arXiv:0907.4798 [hep-th]].

\bibitem{Banks:1998nr}
  T.~Banks and M.~B.~Green,
  ``Non-perturbative effects in AdS(5) x S**5 string theory and d = 4 SUSY
  Yang-Mills,''
  JHEP {\bf 9805} (1998) 002
  [arXiv:hep-th/9804170].

\bibitem{Gubser:1998nz}
  S.~S.~Gubser, I.~R.~Klebanov and A.~A.~Tseytlin,
  ``Coupling constant dependence in the thermodynamics of N = 4  supersymmetric
  Yang-Mills theory,''
  Nucl.\ Phys.\  B {\bf 534} (1998) 202
  [arXiv:hep-th/9805156].

\bibitem{Pawelczyk:1998pb}
  J.~Pawelczyk and S.~Theisen,
  ``AdS(5) x S(5) black hole metric at O(alpha'**3),''
  JHEP {\bf 9809} (1998) 010
  [arXiv:hep-th/9808126].

\bibitem{deHaro:2002vk}
  S.~de Haro, A.~Sinkovics and K.~Skenderis,
  ``A supersymmetric completion of the R**4 term in IIB supergravity,''
  Phys.\ Rev.\  D {\bf 67} (2003) 084010
  [arXiv:hep-th/0210080].

\bibitem{deHaro:2003zd}
  S.~de Haro, A.~Sinkovics and K.~Skenderis,
  ``On alpha' corrections to D-brane solutions,''
  Phys.\ Rev.\  D {\bf 68} (2003) 066001
  [arXiv:hep-th/0302136].

\bibitem{Peeters:2003pv}
  K.~Peeters and A.~Westerberg,
  ``The Ramond-Ramond sector of string theory beyond leading order,''
  Class.\ Quant.\ Grav.\  {\bf 21} (2004) 1643
  [arXiv:hep-th/0307298].

\bibitem{Cadabra1}
  K.~ Peeters.
  ``A field-theory motivated approach to symbolic computer algebra
  cs.SC/0608005,''
  Comp. Phys. Commun. {\bf 176} (2007) 550-558.

\bibitem{Cadabra2}
  K.~ Peeters.
  ``Introducing Cadabra: a symbolic computer algebra system for field theory
  problems,''
  hep-th/0701238.

\bibitem{Arnold:2000dr}
  P.~B.~Arnold, G.~D.~Moore and L.~G.~Yaffe,
  ``Transport coefficients in high temperature gauge theories. 1. Leading log
  results,''
  JHEP {\bf 0011} (2000) 001
  [arXiv:hep-ph/0010177].

\bibitem{Liu}
  H.~Liu, K.~Rajagopal and U.~A.~Wiedemann,
  ``Wilson loops in heavy ion collisions and their calculation in AdS/CFT,''
  JHEP {\bf 0703} (2007) 066
  [arXiv:hep-ph/0612168].

\bibitem{Rajagopal}
  J.~Casalderrey-Solana, H.~Liu, D.~Mateos, K.~Rajagopal and U.~A.~Wiedemann,
  ``Gauge/String Duality, Hot QCD and Heavy Ion Collisions,''
  arXiv:1101.0618 [hep-th].

\bibitem{Aarts:2007wj}
  G.~Aarts, C.~Allton, J.~Foley, S.~Hands and S.~Kim,
  ``Spectral functions at small energies and the electrical conductivity in
  hot, quenched lattice QCD,''
  Phys.\ Rev.\ Lett.\  {\bf 99} (2007) 022002
  [arXiv:hep-lat/0703008].

\bibitem{Ding:2010ga}
  H.~T.~Ding, A.~Francis, O.~Kaczmarek, F.~Karsch, E.~Laermann and W.~Soeldner,
  ``Thermal dilepton rate and electrical conductivity: An analysis of vector
  current correlation functions in quenched lattice QCD,''
  Phys.\ Rev.\  D {\bf 83} (2011) 034504
  [arXiv:1012.4963 [hep-lat]].

\bibitem{Hassanain:2011}
  B.~Hassanain and M.~Schvellinger, work in preparation.




\end{thebibliography}
\end{document}